# Informing Innovation Management: Linking Leading R&D Firms and Emerging Technologies

## Xian Gong*

Faculty of Engineering and Information Technology, University of Technology, Sydney NSW, Australia.
E-mail: xian.gong@student.uts.edu.au
* Corresponding author

## Claire McFarland*

1. Beesness Pty Ltd. Australia
2. Faculty of Engineering and Information Technology, University of Technology, Sydney, NSW, Australia.
E-mail: claire.mcfarland@gmail.com
* Corresponding author

## Paul McCarthy

1. The Data Science Institute, University of Technology, Sydney, NSW, Australia.
2. School of Computer Science and Engineering, UNSW Sydney, NSW, Australia
E-mail: paul@onlinegravity.com

## Colin Griffith

Data61, Commonwealth Scientific and Industrial Research Organisation (CSIRO) Australia.
E-mail: colin.griffith@data61.csiro.au

## Marian-Andrei Rizoiu

Faculty of Engineering and Information Technology, University of Technology, Sydney, NSW, Australia.
E-mail: Marian-Andrei.Rizoiu@uts.edu.au

**Abstract:** Understanding the relationship between emerging technology and research and development has long been of interest to companies, policy makers and researchers. In this paper new sources of data and tools are combined with a novel technique to construct a model linking a defined set of emerging technologies with the global leading R&D spending companies. The result is a new map of this landscape. This map reveals the proximity of technologies and companies in the knowledge embedded in their corresponding Wikipedia profiles, enabling analysis of the closest associations between the companies and emerging technologies. A significant positive correlation for a related set of patent data validates the approach. Finally, a set of Circular Economy Emerging Technologies are matched to their closest leading R&D spending company, prompting future research ideas in broader or narrower application of the model to specific technology themes, company competitor landscapes and national interest concerns.

## 1 Introduction and context

Understanding how emerging technologies are being used by companies, including their capabilities and specialisations, is of increasing interest for a wide range of stakeholders including investors, customers, suppliers, employees, as well as researchers and policymakers.

However, traditional data sources, taxonomies, and methodologies have limitations when it comes to understanding the dynamic and nascent nature of emerging technologies, and the individual capabilities and strengths of the leading firms investing in research and development. For example, patent analysis -- a key source for many existing studies -- works well for some industries such as pharmaceuticals but less well for others such as software development.

The foundation of this research is a methodology to create a dynamic and structured list of the most significant emerging technologies using machine learning trained on the knowledge contained in Wikipedia. This method leverages the output from the tens of thousands of scholars and experts who contribute to articles about technology on Wikipedia, the world's largest and most read reference work in history and assessed as highly reliable for science and technology information (Bruckman, 2022).

The resulting list of 100 emerging technologies (the ET100) has been identified based on analysis of over 50,000 Wikipedia articles most related to the concept of emerging technologies and grouped into hierarchical clusters (that can be interrogated at different levels of detail). The significance of the ET100 list has been demonstrated through positive correlations with the level of attention paid to the Wikipedia articles and current investment in these technologies. Additionally, the ET100 maps to other lists of emerging technologies published by organisation such as the OECD, WEF and Gartner.

A similar machine learning methodology is then used to map the relationship between the ET100 and the activities of the businesses investing the largest sums in R&D worldwide. This approach uses the European Commission's most recent list of the world's top spending R&D companies – the 2022 EU Industrial R&D Scoreboard (Grassano et al 2022), and the knowledge embedded in their corresponding Wikipedia profiles, to find the closest association between the companies and emerging technologies.

Recent research has demonstrated the potential of deriving detailed technical information and other latent knowledge from word embedding language models (Tshitoyan etal, 2019; McFarland etal, 2022). In this paper we illustrate the potential to use the latent knowledge in a novel word embedding language model to create new maps of:

- Leading companies by R&D investment and their relationship to one another
- The global emerging technology landscape
- The relationship between these leading R&D investing companies and an established set of emerging technologies (the ET100).

## 2 What is the problem?

The relationship between business R&D and emerging technologies is difficult to determine at scale and on a dynamic basis.

This is the case for two main reasons, firstly because business R&D is often tightly directed towards the development of new products and services, and secondly the way in which information about the nature of the business R&D is communicated by firms, and collected by national reporting organisations.

The evidence of business R&D focus on the development of products and services can in some cases be gleaned from the record of patents applied for, from company annual reports and media releases. There are some examples of companies being public about their investment in advancing understanding in emerging technologies[1] but much of the detail of business R&D is hidden.

The academic literature is a rich source of business R&D case studies, however it is lacking in linking a generalised group of firms (such as is found in the EU Industrial R&D Investment Scoreboard) with a robust list of emerging technologies.

While national governments are interested in levels of business R&D expenditure (termed BERD in relation to national accounts) because of its contribution to economic growth, in this context there is generally little information collected about the nature of the R&D investment beyond broad categories based either on traditional industry classifications or custom expert derived taxonomies. Business R&D expenditure is measured at a firm level, often reported at an industry or general research field level[2] and compared at a global level by organisations such as the OECD.

By contrast emerging technologies are often examined at an individual technology level or as a category by researchers and policy makers alike. Existing approaches to understanding emerging technologies as a group rely mostly on established legacy taxonomies, such as Standard Industrial Classification codes -- long held to be inadequate in examining high tech environments (Kile and Phillips, 2009); or on expert developed categories that are not dynamic and therefore are hard to maintain. Neither resulting group of emerging technologies provides insights into the complex relationships or associations with other technologies let alone with any business.

There is a gap therefore in looking at the relationship between those firms spending the most on R&D globally and the emerging technology landscape. Since both R&D and emerging technologies play an important role in innovation, better understanding each of these factors and the relationship between them is of key interest to innovation managers.

## 3 Current understanding, what do we know about the problem?

Looking in more detail at the current situation in relation to business R&D and emerging technologies from the perspectives of both commercial interests and policy makers provides important background to this research.

*Importance of emerging tech to companies*

As well as being at the heart of economic growth, technological change is having a profound influence on reshaping the worldwide economy as is evidenced by almost any study of recent history. Looking at publicly listed companies, over the last 30 years the largest companies in the world have changed from predominantly car and fossil fuel manufacturers to digital tech giants.

However, from a company level perspective, technology is so pervasive that it can be hard to determine when a new technology is worth paying attention to. Indeed, the academic definition of emerging technologies as novel, relatively fast growing, with the potential for significant impact on society and the economy while being also somewhat uncertain and ambiguous (Rotolo, Hicks, & Martin, 2015) presents challenges for companies in terms of balancing the risk and reward of deciding to make an investment.

Yet, estimates of the size of the 'New Technologies' market in 2023 exceed $US 1.3 trillion[3.] IDC Global ICT Spending forecasts for 2020 – 2023 predict that over the next 5-10 years new technologies such as robotics, AI and AR/VR will grow to 25% of ICT spending, particularly as firms generate cost savings from implementation of cloud and automation services. These estimates are important inputs for company decision making as they seek to deliver value to shareholders and customers, invest in employees, partner with suppliers, and support the communities in which they operate[4].

An element of this decision making is whether a company will adopt emerging technology, assemble emerging technology into new products and services or develop emerging technology. Elements of business R&D exist within each of these approaches.

The extent to which there are links between leading R&D spending companies and individual emerging technologies is the key focus of this paper.

*Importance of emerging tech to policy makers*

As recognised formally by the United States government in their announcement of the Office of the Special Envoy for Critical and Emerging Technology, "technology is increasingly central to geopolitical competition and the future of national security, economic prosperity, and democracy"[5]. While much of the commentary is inevitably focused on issues of national security and democracy, the emphasis on economic prosperity is of key interest to businesses. The US government has identified a list of critical technologies which broadly encompasses biotechnology, advanced computing, artificial intelligence, and quantum information technology[6].

Likewise, the European Union has recently released a draft report on critical technologies for security and defence[7] and announced a new 'Observatory of Critical

Technologies' with a focus on value and supply chains particularly for space, defence and related civil sectors[8].

These two come together in the US-EU Trade and Technology Council created to co-ordinate key global trade, economic and technology issues with specific focus on a number of emerging technologies[9].

The 2022 EU Industrial R&D Investment Scoreboard, reflects the EU's focus on understanding how Europe compares with both the US and China in terms of overall investment in business R&D; where there are strengths and how the EU wants to excel with 'green and digital transformations'. The New European Innovation Agenda, released in July 2022, has a goal of promoting 'firm creation and growth in emerging technologies to trigger spill overs between sectors'.

Signalling by governments of areas of focus inevitably concentrates investment by businesses as it provides the certainty conducive to a strong investment climate.

*Importance of R&D to firms*

It is clear from the 2022 EU R&D Scoreboard that a significant amount of funding is directed by firms towards R&D. In 2021 the 2,500 companies investing the most in R&D in the world account for approximately 86% of the world's business funded R&D, a total of 1093.9 billion Euro[10].

The relationship of business R&D expenditure and business performance is widely studied by academics and business consultants alike, with evidence that R&D expenditure increases future profitability (Branch 1974; Sougiannis 1994; Eberhart et al. 2004), and has a positive impact on market value (Sougiannis 1994; Armstrong et al. 2006) and stock returns (Lev and Sougiannis 1996; Chan et al. 2001; Kim, 2022).

The relationship of all this business R&D activity with emerging technologies is the key question of interest in this paper.

*Importance of business R&D to policy makers*

R&D investment has long been found to be a key driver of economic growth. Business R&D makes up a significant portion of the Gross Expenditure on R&D for many countries, leading to policy instruments to stimulate business R&D, including both favourable tax treatment and direct subsidies for specific industries. The practice of comparing the levels of R&D investment (public and private) at a national level[11] has become an indicator of commitment and performance (Schot and Steinmueller, 2018). The notion of R&D spillovers, where investment by one entity (for example government) has a beneficial impact to other entities (for example businesses) (Grossman and Helpman, 1992; Griliches, 1992), is a key consideration of government investment in R&D. Additionally, the commercialisation of government R&D by companies has become an area of increasing focus[12].

## 4 Limits of current understanding, thesis and research questions

The macro level intersection of these things is interesting for policy makers, academics, and future-focused companies. Current available data is isolated. For example, media releases / annual report updates from firms about their R&D projects provides a biased (company initiated) perspective. The value of R&D expenditure captured by OECD governments provides a macro view, and while this provides some insight at the industry level, it gives little indication of expenditure direction at the technology level.

Since both R&D at leading firms and the relationship of these firms to emerging technologies play an important role in innovation, better understanding each of these factors and the relationship between them is of key interest to innovation managers.

This paper sets out the thesis that word embedding language models can provide ways to map the proximity of emerging technologies to leading R&D investing companies (and vice versa) and that these maps can be cross validated through established third party data such as patent counts. In doing so it seeks to address the following questions:

- Which emerging technologies are closest to the top 500 firms by R&D expenditure?

- For each of the 100 identified leading emerging technologies, which company from the top 500 R&D spending companies is closest?

- What does this combined dataset look like from a two-dimensional perspective? What are the noticeable clusters? What outliers appear? What further questions does this raise?

## 5 Methodology

This innovation-in-practice paper draws on previous work in emerging technology categorisation using machine learning clustering techniques and links a robust set of defined leading emerging technologies to the firms spending the most on R&D measured at an international level.

The list of emerging technologies (the ET100) has been constructed by mapping thousands of individual technologies and their relationships to each other using machine learning analysis of the similarity between the way different technologies are described. This approach produces a novel multi-level technology taxonomy that reveals the complex relationships between all the technologies, providing insight into the most adjacent technologies and those that are most connected within different technology ecosystems. It also serves as a foundation that can be used to interrogate other sources of data.

The natural language processing tool used in this research has been trained on the full corpus of English language Wikipedia consisting of 4.4 million articles and 1.9 billion terms created by over 40 million editors. Known as Wikipedia2vec (Yamada etal, 2018), this toolkit represents defined entity vectors in the same high-dimensional space. It learns high-quality word embedding from Wikipedia, placing semantically similar words and entities close to one another in the vector space, and incorporating additional information from Wikipedia, such as article titles, links and categories. The entity vectors are particularly powerful because they are trained on a large, multilingual corpus of Wikipedia articles, which covers a wide range of topics and domains. This means that the entity vectors will likely capture a wide variety of semantic relationships between entities, including those that span different languages and cultures.

The research work underpinning this paper defines as entity vectors all the sub-categories and pages contained within the top level category of emerging technologies in Wikipedia; and all the Wikipedia pages associated with the businesses listed in the 2022 EU Industrial R&D Investment Scoreboard. An algorithm was designed to reveal the embeddings or 'similarity' of the defined entities also referred to here as links or associations. The resulting model is referred to from here as the 'ET100/R&D Links' model.

Since technologies and organisations are different types of entities, the concept of orthogonal matrices in machine translation is introduced to rotate the company space to align with the technology space. These matrices are particularly useful because they preserve the distances and angles between vectors, which can help to maintain the semantic relationships between firms and technologies during the alignment process. Dimensionality reduction and clustering analysis creates visualisations that can effectively convey the relationships among technologies and companies. Moreover, it improves interpretability and identifies relationships due to grouping similar entities together into clusters.

The resulting 'ET100/R&D Links' model can be visualised as a map of the 100 identified emerging technologies and the leading 700 or so companies by R&D expenditure. Also revealed is the closest emerging technology to each of these companies, the closest company to each emerging technology and the relationships between individual companies and individual technologies, and between groups of companies and groups of technologies.

The notion of 'similarity' and 'distance' refer to the mathematical translation of the positioning of the emerging technologies and the companies in the 'ET100/R&D Links' model. Each emerging technology and company is represented by a unique vector, enabling the distance between each to be measured.

## 6 Findings

The novel approach to examining the relationships between emerging technologies and leading R&D companies resulting in the 'ET100/R&D Links' model, visualised as a map, enables investigation of a number of exploratory research questions and raises others.

*A map of the emerging technologies and leading R&D companies reveals clusters and white space*

Taking a broad view initially, Figure 1 shows an optimised and cross-validated map of the companies and the emerging technologies. There are clearly several clusters of companies and technologies. There is also a section of the map that is less dense – where there is more white space between companies. Note the size of the Company bubbles reflects the magnitude of their R&D investment. Larger equals more.

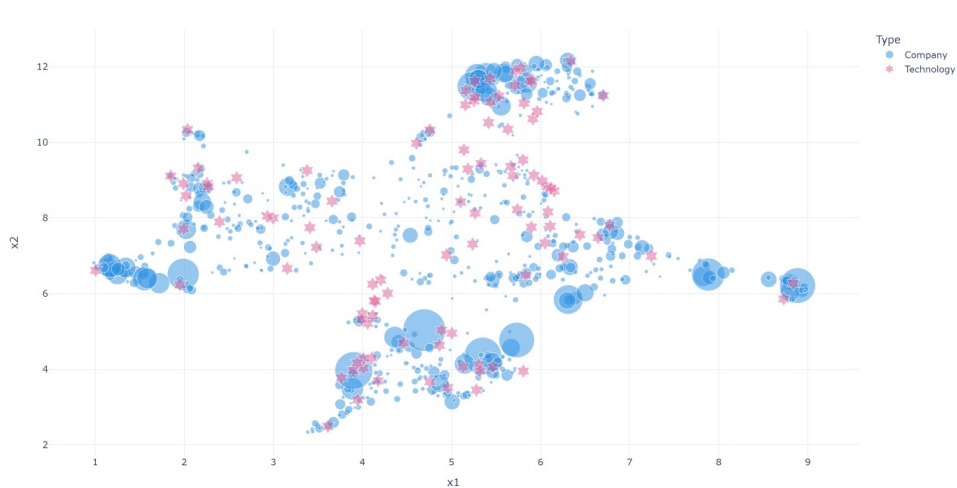

**Figure 1** Visualisation of emerging technologies and leading R&D spending companies map showing distances and relative sizes of R&D expenditure.

Nine natural groups or clusters of companies and technologies have been identified within this map using hierarchical clustering, as shown in figure 2. GPT3 was used to suggest labels to describe each group of companies and the authors have settled on the current list taking this input and the most distinctive industry in the cluster into account.

Two company clusters have high concentrations of close emerging technologies – the Technology and Financial Services cluster (17 technologies) and the Pharmaceuticals and Biotech cluster (20 technologies). This is unsurprising and reflects the maturity and widespread application of both these fields.

By contrast, the company cluster for Telecommunications has the least concentration of close emerging technologies, with just three emerging technologies close to the companies within this clusters. Additionally while the closest adjacent cluster is Electronics and Manufacturing, the Telecommunications cluster is some distance away from several of the emerging technologies often associated with the telecommunications industry, such as Local Positioning System and Internet-of-Things.

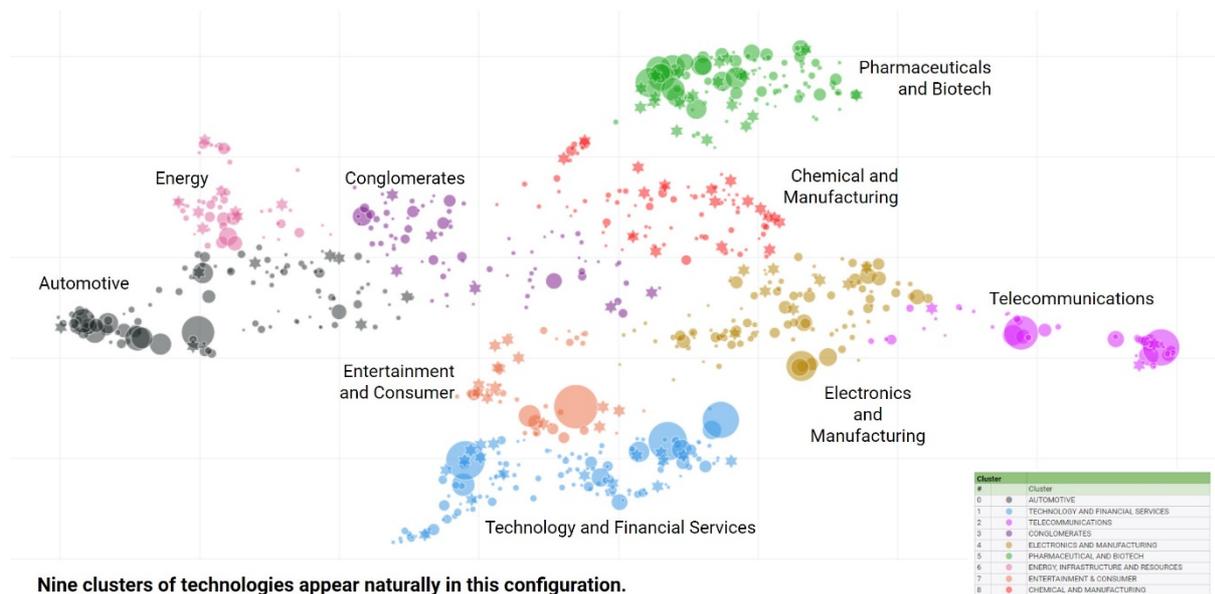

Nine clusters of technologies appear naturally in this configuration.

**Figure 2** Hierarchical clustering reveals nine natural groups of leading R&D spending companies and emerging technologies.

Returning to the area of relative white space within the map, where there is less density of both companies and technologies, there are two noticeable insights. Firstly, the technologies within this location tend to be more specific in terms of both definition and application than some of the technologies appearing in the more closely linked clusters. For example, Small Satellite and Distributed Acoustic Sensing have very particular applications at this point in time.

The second insight is the companies within this white space are generally spending less on R&D comparatively, with a large proportion below 500m Euro. There are no 'mega' spending companies within this part of the map. Note that the more tightly defined clusters of companies and technologies tend to contain a number of 'mega' spenders and a number of closely linked technologies. Could this be evidence of the development of R&D ecosystems within these clusters?

*Zoomed in perspective one: Closest emerging technologies to the leading R&D spending companies*

Beyond the map visualisation it is possible to zoom in to the 'ET100/R&D Links' model to examine two perspectives. Firstly, the closest technologies to the companies appearing in the 2022 EU Industrial R&D Investment Scoreboard. While revealing the closest technologies, this also provides a comparative perspective on the relative position of the set of closest technologies to different companies. For some companies there are a number of technologies clustered together, for others there is a single closest technology and then some distance to the next closest technology, for others there is a more even spacing between close technologies. Figure 3 sets out one subset of companies with a variety of top five closest technology positions to illustrate this.

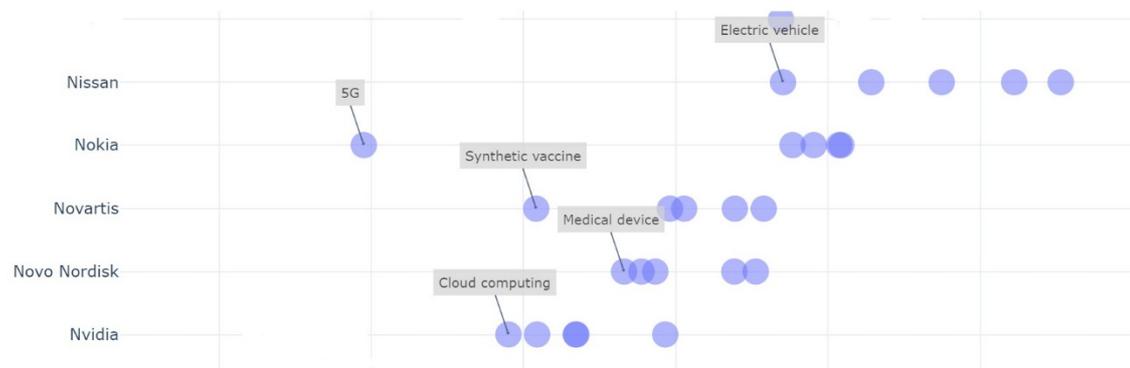

**Figure 3** A subset of leading R&D spending companies with their closest emerging technologies.

Looking in detail at Figure 3, it is clear that there is some distance between 5G, the closest technology to telecommunications company Nokia, and the next closest technologies which are Unified Communications, Internet of Things, Local Positioning System and Cloud Computing, with these latter four technologies clumped together.

By contrast, the closest technology to automotive company Nissan is Electric Vehicle, although this is only about as close as Unified Communications is to Nokia. The other close technologies to Nissan are Fleet Management, Fuel Cell, Autonomous car and Energy content of biofuel. These four technologies are relatively evenly spaced and all at a further distance from Nissan than the position of the top five close technologies of Nokia.

Novartis and Novo Nordisk are both pharmaceutical companies and while their closest technologies are different, with Synthetic vaccine for Novartis and Medical device for Novo Nordisk, they do share some similarities beyond this. For both companies, their second and third closest technologies are Biotechnology followed by Cell Therapy, both of which are relatively closer to Novo Nordisk. Beyond this for Novartis, Vaccine Efficiency followed by Neurotechnology round out the top five closest technologies, while these positions are held by Regenerative Medicine and Clinical Oncology for Novo Nordisk.

*Zoomed in perspective two: Closest companies to emerging technologies defined in ET100*

Looking at the alternative perspective of which companies in the EU Industrial R&D Investment Scoreboard are closest to the emerging technologies within the ET100 also shows some strong variance in positioning in a one dimensional view.
Figure 4 shows a subset of technologies to illustrate this.

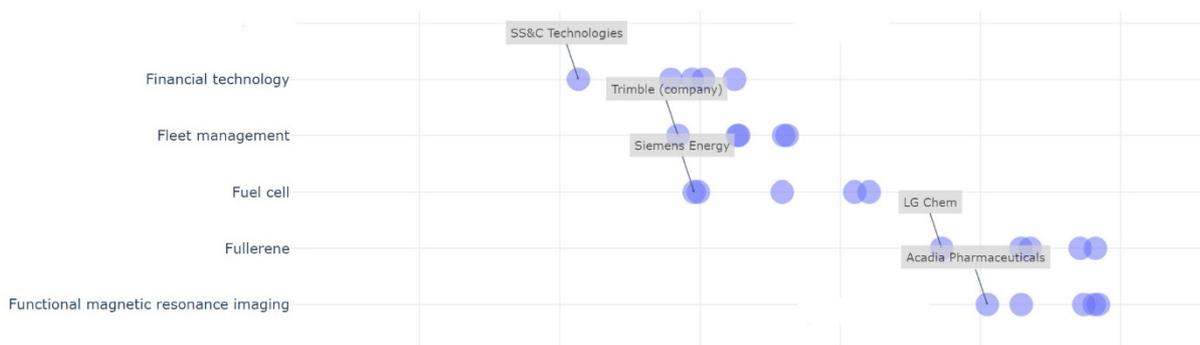

**Figure 4** A subset of emerging technologies with their closest leading R&D spending companies

In this group of technologies Fullerene and Functional magnetic resonance imaging have no companies relatively close to them suggesting they are still nascent; whereas the more mature financial technology (fintech), fleet management and fuel cell technologies have established companies close to each of them.

Across the whole set of 100 emerging technologies, some companies appear to have more of a dominant lead in their association with technologies when compared to nearest rivals. Illumina for example in Whole Genome Sequencing is a long way ahead of its nearest rival, and for 5G a number of global telcos and OEMs led by China Mobile, Nokia and CommScope are in relatively close proximity. This raises the question about whether the closeness of association corresponds to relative market share or other measures of advantage, which is perhaps an opportunity for future research.

*Zoomed in perspective three: Circular Economy Technologies and leading R&D spending companies*

Looking specifically at Circular Economy Technologies (CETs), the 'ET100/R&D Links' model can be used to investigate the closest emerging technologies to the top companies identified as the leading inventors in Circular Economy Technologies (CETs) set out in the 2022 EU Industrial R&D Investment Scoreboard [13].

For example, European multinational BASF is identified in the 2022 EU Industrial R&D Investment Scoreboard as being a top European company inventing in CETs in multiple industries (Construction, Chemicals and Plastics, Fertilisers, Metals, and Batteries & Fuel Cells). The five closest emerging technologies to BASF in the 'ET100/R&D Links' model are Smart Material, Biotechnology, Synthetic Vaccine, Semiconductor, and Carbon capture and storage.

Conversely, the 'ET100/R&D Links' model can also be interrogated from the emerging technology perspective. Taking a handful of emerging technologies with application to the Circular Economy and determining which company from the 2022 EU Industrial R&D Investment Scoreboard is closest as set out in Table 1 provides an insight into the CET landscape.

**Table 1** Identified Circular Economy Emerging Technologies matched to their closest leading R&D spending company within the 'ET100/R&D' Links model

| *Circular Economy Emerging Technology* | *Closest Company from the 2022 EU Industrial R&D Investment Scoreboard* |
|---|---|
| Biocatalysis | Novozymes |
| Biomaterial | Mitsui Chemicals |
| Carbon capture and storage | Siemens Energy |
| Energy management | Huaneng Power |
| Grid energy storage | Goldwind |
| Microgrid | Siemens Energy |
| Precision agriculture | Hexagon AB |

Source: ET100/R&D Links model

*Patent analysis validation*

Finally, since this is exploratory research, an additional question of interest is whether the relationship between this set of leading R&D spending companies and emerging technologies as constructed in the 'ET100/R&D Links' model is consistent with existing research or third-party data? To determine this an analysis of global patent data as it relates to the emerging technologies and the companies in the 'ET100/R&D Links' model was undertaken.

Looking at the counts of global patents for all emerging technologies in the ET100 for a set of top companies from the EU Industrial R&D Investment Scoreboard there is a significant positive correlation for almost half (n=48) with the similarity measures within the 'ET100/R&D Links' model.

The medical technologies within the ET100, where there is a lot of patenting activity (Clinical Oncology, Vaccine efficacy, Biotechnology) are among the most strongly correlated. Many of the technologies that are not significantly correlated are either very new and may not have much patent momentum or are very specific technologies, which also may not have much patenting activity. Recall also that patents are highly applicable in some industries (eg pharmaceuticals) and much less so in others (software). This raises another opportunity for future research to better understand the significance of this pattern of correlation.

An example of the patent analysis undertaken is shown in Figure 5, which sets out the correlation of patent counts to 'ET100/R&D Links' model similarity measures for Speech Recognition.

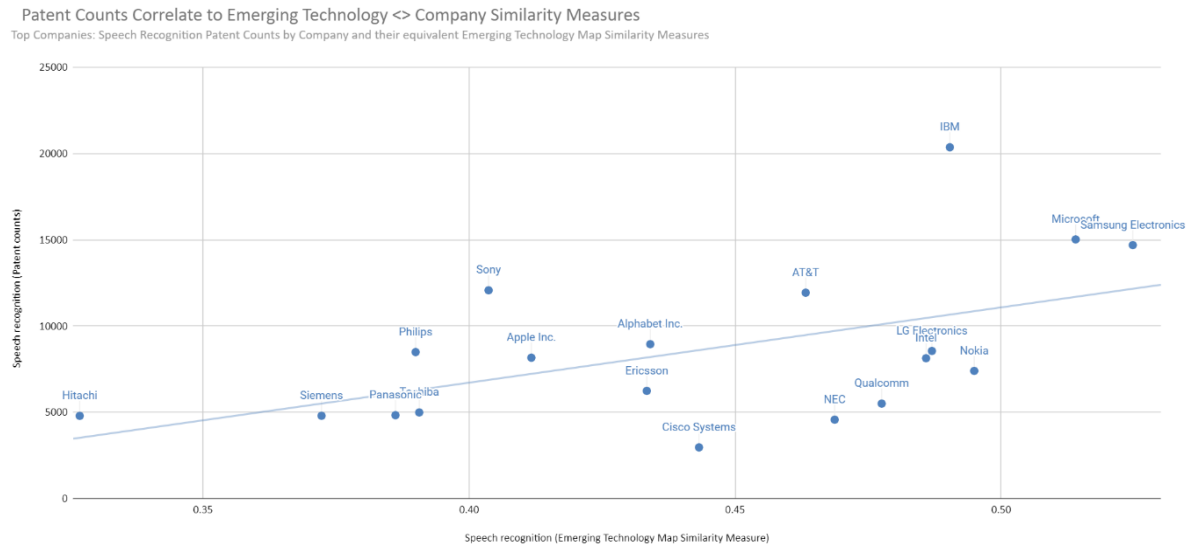

**Figure 5** Example of significant positive correlation between patent counts and 'ET100/R&D Links' model for Speech Recognition technology.

## 7 Limitations, significance (and potential new areas of research)

In taking this approach to investigating the linkages between the top R&D spending companies and a defined list of emerging technologies some qualifications are required, and this is to do with the characterisation of the linkages for some companies with some of the emerging technologies. Due to the nature of the methodology, the exact nature of the association between a technology and a company may require further interpretation. It could show an existing investment in a technology, or an adjacency that could reveal a potential to leverage a technology in a novel manner, or it could show other associations related to some specific use or interest in a technology.

It is not certain that the linkages observable in the 'ET100/R&D Links' model are due to investments by the companies in question into the emerging technology in question. For example, a company such as food multinational Mondelez appears to be associated with vaccine efficacy due to a policy to vaccinate all its employees rather than involvement in the development of such technology in the vaccine space. Additionally, some linkages reflect multiple or ambiguous word meanings. For example, the term microsatellite has association with both genomic analysis within the biotechnology field, and the space industry.

This variation is a natural outcome from a natural language processing model such as the 'ET100/R&D Links' model underpinning this paper and is the reason why expert interpretation is so important in research. In this case to interpret the nature of the linkages between specific companies and technologies based on specific context. To ask the question "why has the embedded knowledge in both these sets revealed a specific association?".

For commercial innovation managers this research provides an insight into the competitive landscape amongst leading R&D firms. This provides rich information for the development of strategic planning and potentially for strategic partnership opportunities or M&A activity. The correlation between the 'ET100/R&D Links' model and the patenting activity of companies and specific technologies (per the Speech Recognition example) reveals that in part, the model reflects technology strength and / or capability. Therefore, the information presented in the 'ET100/R&D Links' model provides insight into company-specific areas of potential investment in emerging technology development.

For policy-makers, insight into the proximity of particular emerging technologies to particular companies may provide another layer of detail as to where R&D investment is flowing, knowledge that may be valuable in a variety of policy contexts.

For innovation researchers this paper and accompanying presentation may be thought-provoking for new avenues of research. For other researchers this may provide ideas for research partnerships.

Beyond the approach examined in this paper, the methodology underpinning the 'ET100/R&D Links' model can be applied more broadly or more narrowly to investigate more extensive lists of technologies and companies. Some examples of potential deep dives into specific technology themes include circular economy technologies, pandemic resilience and response related technologies, quantum computing related technologies. Or the methodology could be applied to examine the competitor landscape by defining a set of competitive companies with a view to developing understanding of the technology landscape within which they operate. This could be insightful for both established companies and new entrants. The methodology could also be applied to custom lists of technologies and companies to examine specific areas of national interest, for example the critical technology situation or the major supply chain environment.

The overall advantage of the methodology is the ability to identify linkages and therefore associations at scale between emerging technologies and leading R&D spending companies. The complexity of the web of linkages is significant and in aggregate accurate. It enables adjacencies to be discovered and explored, at a company level, at an emerging technology level, and at a national/regional level. This research opens up many additional questions, some of which have been set out here. The authors are keen to explore further opportunities arising from the ISPIM 2023 conference where this work is being presented for the first time.

---

[1] For example, a number of big tech companies, including IBM, Microsoft, Google and Intel are making significant investments in quantum computing R&D via VentureBeat https://venturebeat.com/data-infrastructure/quantum-progress-how-ibm-microsoft-google-and-intel-compare/ accessed Apr 2023.

[2] Examples of Fields of Research collected include 'Information and Computing Sciences', 'Engineering', and 'Agricultural, Veterinary and Food Sciences'.

[3] IDC – Global ICT Spending Forecast 2020 – 2023

[4] Statement on the purpose of a corporation via Business Roundtable https://system.businessroundtable.org/app/uploads/sites/5/2023/02/WSJ_BRT_POC_Ad.pdf accessed Apr 2023

[5] Department Press Briefing - January 3, 2023 via United States Department of State https://www.state.gov/briefings/department-press-briefing-january-3-2023/ accessed Apr 2023

[6] Critical and Emerging Technologies List Update via Whitehouse.gov https://www.whitehouse.gov/wp-content/uploads/2022/02/02-2022-Critical-and-Emerging-Technologies-List-Update.pdf. Also US Critical and Emerging Technology Strategy via Vivekananda International Foundation (vifindia.org) https://www.vifindia.org/article/2023/january/30/us-critical-and-emerging-technology-strategy accessed Apr 2023

[7] Critical technologies for security and defence: state of play and future challenges (2022/2079 (INI)) https://www.europarl.europa.eu/doceo/document/ITRE-PR-738598_EN.pdf via European Parliament Committee of Industry, Research and Energy accessed Apr 2023.

[8] Note, the EU approach is a new endeavour due to a new Action Plan on synergies between civil, defence and space industries. https://ec.europa.eu/commission/presscorner/detail/pt/QANDA_21_652 Q&A: Synergies between civil, defence and space industries via Europa.eu accessed Apr 2023.

[9] U.S.-EU Trade and Technology Council https://www.trade.gov/useuttc via International Trade Administration accessed Apr 2023.

[10] The 2022 EU Industrial R&D Investment Scoreboard | IRI (europa.eu) https://iri.jrc.ec.europa.eu/scoreboard/2022-eu-industrial-rd-investment-scoreboard via Europa.eu accessed Apr 2023.

[11] The OECD collects and publishes this data in multiple forms on a regular basis. OECD publications include the Main Indicators and Science and Technology (MSTI), the OECD Science Technology and Industry Scoreboard and numerous datasets. https://www.oecd.org/sti/inno/researchanddevelopmentstatisticsrds.htm via Research and Development Statistics (RDS) – OECD accessed Apr 2023

[12] For example, the United States government has a series of initiatives related to commercialising federally funded R&D, including https://www.nist.gov/tpo/lab-market via Lab-to-Market (L2M) | NIST. In Australia, the NSW Government also has a strong focus, https://www.dpc.nsw.gov.au/publications/categories/turning-ideas-into-jobs-accelerating-research-and-development-in-nsw/ Turning ideas into jobs: Accelerating research and development in NSW via Premier & Cabinet accessed Apr 2023.

[13] Table 5.3 Top five Scoreboard companies inventing in CETs per industry (2010-2019) from the 2022 EU Industrial R&D Investment Scoreboard | IRI (europa.eu) https://iri.jrc.ec.europa.eu/scoreboard/2022-eu-industrial-rd-investment-scoreboard via Europa.eu sets out companies from both the Global and European perspective. A small number of examples are used here to illustrate the strength and applicability of the 'ET100/R&D Links' model.